\documentclass{appolb}
\usepackage{graphicx}
\usepackage{epsfig,epsf}
\usepackage{amsbsy}
\usepackage{amsfonts}
\usepackage{textcomp}
\usepackage{amssymb}
\usepackage{mathrsfs}
\usepackage{color}

\begin{document}
\title{The true  face of quantum decay processes: \\Unstable systems in rest and in motion
\thanks{Presented at $2^{nd}$ Jagiellonian Symposium on Fundamental and Applied Subatomic Physics,
June 4 --- 9, 2017, Krak\'{o}w, Poland }%
}
\author{Krzysztof Urbanowski\footnote{e--mail:  K.Urbanowski@if.uz.zgora.pl, k.a.urbanowski@gmail.com}
\address{
University of Zielona G\'{o}ra, Institute of Physics, \\
ul. Prof. Z. Szafrana  4a, 65--516 Zielona G\'{o}ra, Poland}
}
\maketitle
\begin{abstract}
We analyze properties of unstable systems at rest and in motion.
\end{abstract}

\section{Introduction}
Since the discovery of the radioactive decay law  by Rutherford and Sody
the belief that the decay law has the exponential form has become common.
This conviction was upheld by Wesisskopf--Wigner theory of spontaneous emission \cite{WW}.
Further studies of the quantum decay process showed that basic principles of the quantum theory
led to
rather widespread belief that a universal feature of the quantum decay process is the presence of three
time regimes of the decay process:
The early time (initial), exponential (or "canonical"), and late time having inverse--power law form \cite{peshkin}.
The question arises, if indeed
this is the true picture of quantum decay processes.

From the standard, text book  considerations one finds that if
the decay law of the unstable particle at rest
 has the exponential form
${\cal P}_{0}(t) = \exp\,[- \frac{{\it\Gamma}_{0}\,t}{\hbar}]$,
then the decay law of the moving particle looks as follows:
\begin{equation}
{\cal P}_{p}(t)
\,= \,\exp\,[-\,\frac{{\it\Gamma}_{0}\,t}{\hbar \,\gamma}], \label{P-p-(t)}
\end{equation}
where $t$ denotes time,
${\it\Gamma}_{0}$ is the decay rate (time $t$ and ${\it\Gamma}_{0}$
are measured in the rest reference frame of the particle)
and $\gamma$ is the
relativistic Lorentz factor.
 Formula (\ref{P-p-(t)}) is the classical physics relation.
It is almost common belief that this formula is valid also for any $t$
in the case of quantum decay processes
and does not depend on the model of the unstable particles considered.
The problem seems to be extremely important
because from some theoretical studies it follows that in the case of quantum
decay processes this relation is valid to a sufficient accuracy only for not
more than a few lifetimes
 $\tau_{0} = \hbar / {\it\Gamma}_{0}$ \cite{stefanovich,shirkov,exner,ku-2014}.
 All the above problems will be analyzed in the next parts of this paper.

\section{Unstable states in the rest system}

The main information about properties of  quantum unstable   systems
is contained in their decay law, that is in their survival probability.
If one knows that the system in the rest frame is in the initial unstable
state $|\phi\rangle \in {\cal H}$, (${\cal H}$ is
 the Hilbert space of states of the considered system), which was prepared at the initial instant $t_{0} =0$,
one can calculate
its survival probability (the decay law), ${\cal P}_{0}(t)$, of the unstable state $|\phi\rangle$ decaying
in vacuum, which equals
\begin{equation}
{\cal P}_{0}(t) = |a_{0}(t)|^{2}, \label{P(t)}
\end{equation}
where $a_{0}(t)$ is  the probability amplitude of finding the system at the
time $t$ in the rest frame
in the initial unstable state $|\phi\rangle$,
\begin{equation}
a_{0}(t) = \langle \phi|\phi (t) \rangle \equiv \langle \phi|\exp\,[-itH]|\phi\rangle, \label{a(t)}
\end{equation}
$H$ is the selfadjoint Hamiltonian of the system considered and $|\phi (t)\rangle$ is the solution of the Schr\"{o}dinger equation
for the initial condition  $|\phi (0) \rangle = |\phi\rangle$. Here the system units $\hbar = c = 1$ is used.
From basic principles of the quantum theory it follows that the amplitude $a_{0}(t)$ can be represented by the Fourier transform of the mass (energy) distribution function $\omega(m)$ as follows   {\cite{khalfin,fonda,fock}:
\begin{equation}
a_{0}(t)  \equiv \int_{\mu_{0}}^{\infty} \omega(\mu)\;
e^{\textstyle{-\,i\,\mu\,t}}\,d{\mu},
\label{a-spec}
\end{equation}
where   $\omega(\mu) \geq 0$ for $\mu \geq \mu_{0}$ and $\omega(\mu) = 0$ for $\mu < \mu_{0}$.

The simplest way to compare the decay law ${\cal P}_{0}(t)$ with the exponential (canonical) decay law
${\cal P}_{c}(t) = |a_{c}(t)|^{2}$, where
$a_{c}(t) = \exp\,[-i\frac{t}{\hbar}(m_{\phi} - \frac{i}{2}{\it\Gamma}_{\phi}]$,
and $m_{\phi}$ is the rest mass of the particle $\phi$ and  ${\it\Gamma}_{\phi}$ is its decay width, is to analyze properties of the following function:
\begin{equation}
\zeta (t) \stackrel{\rm def}{=} \frac{a_{0}(t)}{a_{c}(t)}. \label{zeta}
\end{equation}
There is
$|\zeta (t)|^{2} = \frac{{\cal P}_{0}(t)}{{\cal P}_{c}(t)}$.
Analysis of properties of this function allows one to visualize all the more subtle differences between ${\cal P}_{0}(t)$ and
${\cal P}_{c}(t)$.

\section{Numerical studies: The Breit--Wigner model}

Results of studies of numerous  models presented in the literature show that
decay curves obtained for these models
are very similar in form to the curves calculated for
 $\omega (\mu)$ having a Breit--Wigner form
 $\omega (\mu) \equiv \omega_{BW} (\mu)$
(see \cite{nowakowski} and analysis in \cite{fonda}):
\begin{equation}
\omega_{BW}(\mu) =  \frac{N}{2\pi}\,  \it\Theta (\mu - \mu_{0}) \
\frac{{\it\Gamma}_{0}}{(\mu-m_{0})^{2} +
(\frac{{\it\Gamma}_{0}}{2})^{2}}, \label{omega-BW}
\end{equation}
where $N$ is a normalization constant and ${\it\Theta}(\mu)$ is a step function.
So to find the most typical properties of the decay curve it is sufficient to make the relevant calculations for  $\omega (\mu)$ modeled by the the Breit--Wigner
distribution of the mass (energy) density $\omega_{BW}(\mu)$.
The typical  form of the survival probability ${\cal P}_{0}(t)$
is presented in Fig (\ref{f1}).
The form of the decay curves depend on the ratio $s_{R} = \frac{m_{R}}{{\it\Gamma}_{0}}$, where  $m_{R} = m_{0} - \mu_{0}$: The smaller
$s_{R}$, the shorter  time when the late time deviations from the exponential form of ${\cal P}_{0}(t)$ begin to dominate.
\begin{figure}[h!]
\begin{center}
\includegraphics[width=70mm]{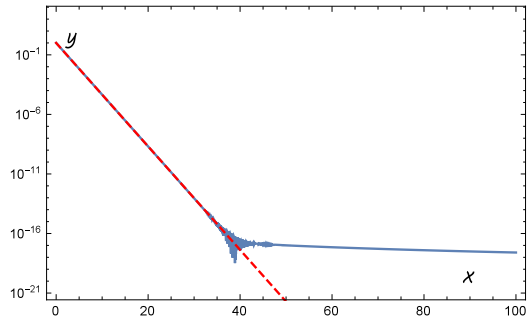}\\
\caption{ Decay curves obtained for $\omega_{BW}(E)$.
Axes: $x =t / \tau_{0} $; $y$: ${\cal P}_{0}(t) = |a_{0}(t)|^{2}$ (The solid line), ${\cal P}_{c}(t) = |a_{c}(t)|^{2}$ (The dotted line). The case $s_{R} = \frac{m_{R}}{{\it\Gamma}_{0}} = 1000$.}
  \label{f1}
\end{center}
\end{figure}
Within the considered model the standard canonical form of the survival amplitude $a_{c}(t)$, is given by the following relation,
$a_{c}(t) = \exp\,[{-i\frac{t}{\hbar}\,(m_{0} - \frac{i}{2}\,{\it\Gamma}_{0})}]$.
${\it\Gamma}_{0}$ is the decay rate and $\frac{\hbar}{{\it\Gamma}_{0}} \equiv \frac{1}{{\it\Gamma}_{0}}= \tau_{0}$ is the lifetime  within the assumed system of units $\hbar = c =1 $ (time $t$ and ${\it\Gamma}_{0}$
are measured in the rest reference frame of the particle).
The typical form of $|\zeta (t)|^{2}$ is presented in Fig (\ref{f2}).

\begin{figure}[h!]
\begin{center}
\includegraphics[width=58mm]{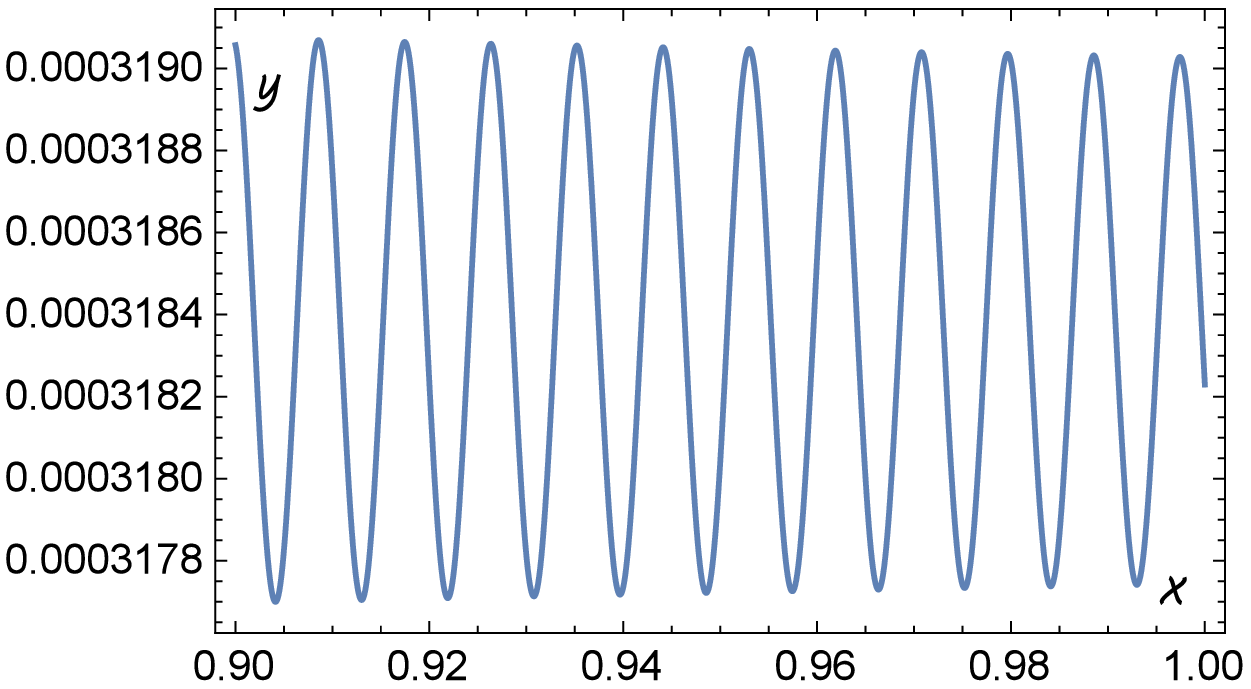}$\;\;\;\;$
\includegraphics[width=58mm]{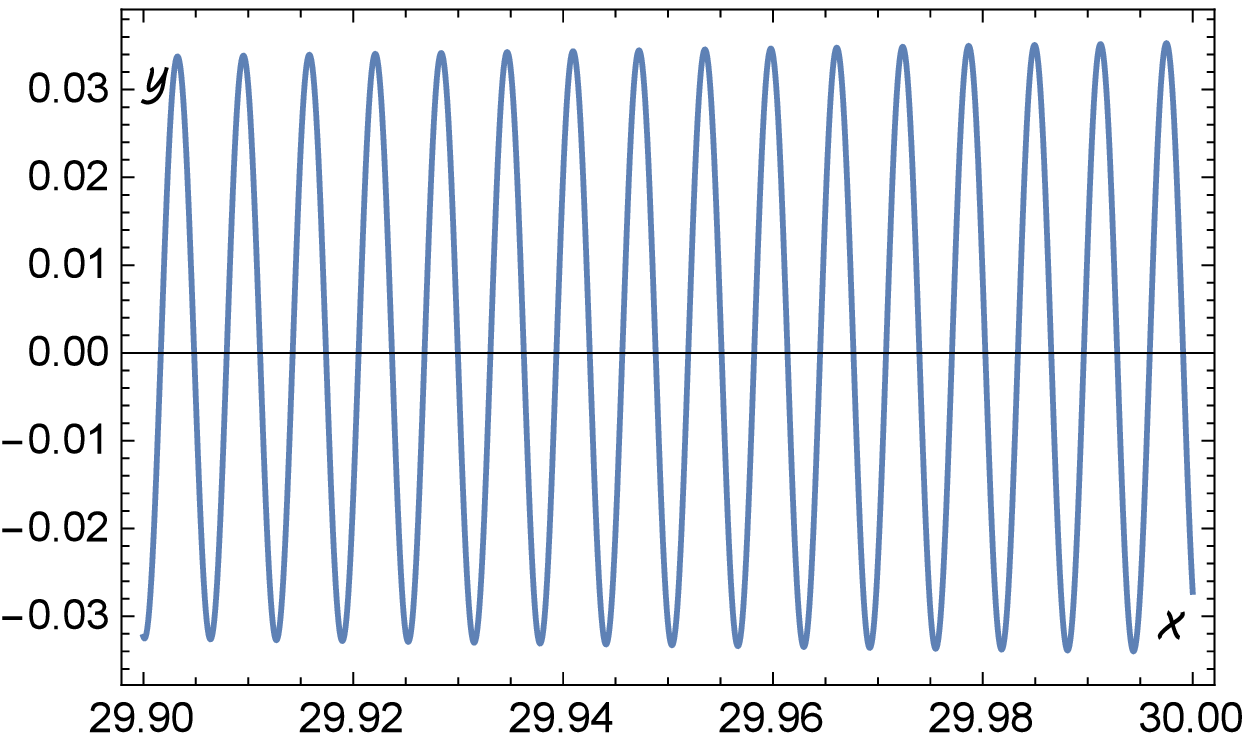}\\
\caption{ A comparison of decay curves obtained for  $\omega_{BW}(\mu)$ with canonical decay curves.
Axes:  $x =t / \tau_{0} $,    $\;\;y$: The function  $f(t) = (|\zeta (t)|^{2} -1) = \frac{{\cal P}_{0}(t)}{{\cal P}_{c}(t)}\, - \,1$, (${\cal P}_{0}(t) = |a_{0}(t)|^{2}$,
 ${\cal P}_{c}(t) = |a_{c}(t)|^{2}$). The case   $s_{R} = 1000$. }
  \label{f2}
\end{center}
\end{figure}
From results of the model calculations presented in Fig (\ref{f2})
it follows that at the initial stage of the "exponential" (or "canonical") decay regime
the amplitude of these oscillations  may be much less than the accuracy of detectors. Then  with increasing time  the amplitude of oscillations grows,
which increases the chances of observing them.
This is a true quantum picture of the decay process at the so--called "exponential" regime of times.

\section{Moving unstable systems}

Analyzing moving unstable systems one can follow the classical physics results and to assume that the unstable systems move with the constant velocity $\vec{v}$, or guided by conservations laws to assume that the momentum $\vec{p}$ of the moving unstable system is constant in time.
The assumption $\vec{v} = const$ was used, eg. by Exner \cite{exner} and also by Alavi and Giunti \cite{carlo}.
Exner obtained result that coincides with the classical result ${\cal P}_{v}(t) \simeq {\cal P}_{0}(t/\gamma)$ but detailed analysis shows that this results was obtained assuming that the velocity $\vec{v}$ is very small. Alavi and Giunti use this assumption and claims that their result is the general one but more detailed analysis of their considerations shows that their conclusion can not be true.
They use the definition (\ref{P(t)})
 of the survival probability
mentioned earlier:
$P_{0}(t) = |a_{0}(t)|^{2}$
of the unstable system in rest.
 The final result is obtained in \cite{carlo} for states
connected with the {\em "reference frame in which the
system is in motion with velocity $\vec{v}$"}. In this new reference
frame the momentum of the particle equals $\vec{k}_{m}$ and $\vec{k}_{m} \neq \vec{p}$,
where $\vec{p}$ is the momentum of the same particle but in the rest frame
of the observer.
The state of the moving unstable particle is described by
a vector $|\Phi_{\vec{v}}\rangle$   which should be an element of
the Hilbert space ${\cal H}_{v}$
connected with this new reference frame in which the system is in motion
but this problem is not explained in \cite{carlo}.
Using states $|\Phi_{\vec{v}}\rangle$ authors of \cite{carlo}
define the amplitude (see (21) in \cite{carlo}),
$a_{\vec{v}}(t;\vec{x}) = \langle \Phi_{\vec{v}}|\exp\,[-itH +
i\vec{P}\cdot \vec{x}]|\Phi_{\vec{v}}\rangle$,
where $\vec{x}$ is a coordinate and $\vec{P}$ is the momentum operator.
The interpretation of the amplitude $a_{\vec{v}}(t;\vec{x})$ is unclear:
The vector
$\exp\,[-itH + i \vec{P}\cdot \vec{x}]\,|\Phi_{\vec{v}}\rangle$ does not
solve the Schr\"{o}dinger evolution equation for the initial condition $|\Phi_{\vec{v}}\rangle$.

Searching  for the properties of the amplitude  $a_{\vec{v}}(t; \vec{x})$
authors of \cite{carlo} use
the integral representation of $a_{\vec{v}}(t; \vec{x})$ as the
Fourier transform of the energy or,
equivalently mass distribution function $\omega (m)$
(see, eg. \cite{khalfin,fonda}) and obtain that (see (39) in \cite{carlo})
\begin{eqnarray}
a_{\vec{v}}(t;\vec{x}) =
\int \, dm\, \Big[\omega(m)
\int\,d^{3}\vec{p}\,|\phi (\vec{p})|^{2}\,e^{\textstyle{-iE_{m}(\vec{k}_{m})\,t\,+\,i\vec{k}_{m}\cdot \vec{x}}}\;\Big], \label{Avx2}
\end{eqnarray}
where $\omega (m) = |\rho (m)|^{2}$ and $\rho (m)$ are the expansion
coefficients of $|\Phi_{\vec{v}}\rangle$ in the basis of eigenvectors
$|E_{m}(\vec{k}_{m}), \vec{k}_{m},m\rangle$
for the Hamiltonian $H$
(see (37) in \cite{carlo}).  $\phi (\vec{p})$ is the momentum distribution
such that $\int d^{3}\vec{p}\,|\phi (\vec{p})|^{2} =1$. The energy
$E_{m}(\vec{k}_{m})$ and momentum $\vec{k}_{m}$ in the new reference
frame mentioned are connected with $E_{m}(\vec{p})$ and $\vec{p}$ in the
rest frame by Lorentz transformations (see (33) --- (35) in \cite{carlo}),
\begin{eqnarray}
E_{m}(\vec{k}_{m}) = \gamma (E_{m}(\vec{p}) + v \, p_{\parallel}),\;\;\;
k_{m\,\parallel} =  \gamma (p_{\parallel} + v E_{m}(\vec{p}),
\label{Ekm-a}
\end{eqnarray}
and $\vec{k}_{m\,\perp} = \vec{p}_{\perp}$,
where $k_{m\,\parallel} \; (\vec{k}_{m\,\perp})$ and
$p_{\parallel}\; (\vec{p}_{\perp})$ are components of
$\vec{k}_{m}$ and $\vec{p}$ parallel (orthogonal)
to the velocity $\vec{v}$, and
$E_{m}(\vec{p}) = \sqrt{m^{2} + \vec{p}^{\;2}}$.

Using the amplitude $a_{\vec{v}}(t;\vec{x})$ authors of \cite{carlo}
define the survival probability ${\cal P}_{\vec{v}}(t)$ of the moving
relativistic unstable particle as (see (40) in \cite{carlo}):
\begin{equation}
{\cal P}_{\vec{v}}(t) = \frac{\int\,d^{3}x\,|a_{\vec{v}}(t,\vec{x})|^{2}}{\int\,d^{3}x\,|a_{\vec{v}}(t=0,\vec{x})}.
\label{Pv}
\end{equation}
then they present main steps of calculations of this probability. In
conclusion they claim that the result of performed calculations
shows that
\begin{equation}
{\cal P}_{\vec{v}}(t) = |a_{0}(t/\gamma)|^{2} \equiv {\cal P}_{0}(t/\gamma), \label{Pv=P0}
\end{equation}
where $\gamma = 1/\sqrt{1 - v^{2}}$  within the system of units used.

To proof this last relation authors of \cite{carlo} limited their considerations to the case when
for the decay width ${\it\Gamma}$, for  mass of the particle $M$ and for the  momentum
uncertainty $\sigma_{p}^{2} = \int \, d^{3}\vec{p}\,|\phi (\vec{p})|^{2}(p_{i})^{2}$, ($i=1,2,3$),
the condition ${\it\Gamma}  \ll \sigma_{p} \ll M$
is assumed to hold. This is crucial condition which allowed them
to approximate the energy $E_{m}(p)$ for all $m$ from the spectrum of $H$
as follows
\begin{equation}
E_{m}(\vec{p})  \simeq m, \label{Emp-1}
\end{equation}
neglecting terms of order $\vec{p}^{\;2}/m^{2}$. Note that integral
(\ref{Avx2}) is taken over all  $m$ from the spectrum
$\sigma (H)$ of $H$. This means that approximation
(\ref{Emp-1}) has to hold for every $m \in \sigma(H)$.
The approximation (\ref{Emp-1}) was used in \cite{carlo} to replace relations
(\ref{Ekm-a}) by the following approximate one,
\begin{eqnarray}
E_{m}(\vec{k}_{m}) &\equiv  & \gamma (E_{m}(\vec{p}) + v \, p_{\parallel})
 \simeq \gamma (m + v \, p_{\parallel}), \label{Ekm-a1}\\
k_{m\,\parallel} &\equiv & \gamma (p_{\parallel} + v E_{m}(\vec{p}) )
\simeq \gamma (p_{\parallel} + v m).
\label{Ekm-a2}
\end{eqnarray}

A discussion
of the admissibility of the mentioned conditions and approximations
uses arguments similar to those one can find, e.g. in \cite{exner}.
The difference is that in \cite{exner} the approximation
$E_{p}(m) \simeq m + \vec{p}^{\;2}/2m$ is used instead of (\ref{Emp-1}).

Finally replacing  $E_{m}(\vec{k}_{m})$ and  $\vec{k}_{m}$ under the integral sign
in (\ref{Avx2}) by  (\ref{Ekm-a1}) respectively (or in \cite{carlo},
in (41) by (33) and (34))
after some algebra authors of \cite{carlo} obtain their  relation (46) that was needed, that is
the  relation denoted as (\ref{Pv=P0}) in this Section.
This result obtained within the conditions and approximations described
above was the basis of the all conclusions presented in \cite{carlo}.

Unfortunately, in \cite{carlo}
there is not any analysis
of physical consequences
of assumed conditions and approximations used. Note that
 \begin{equation}
(\, E_{m}(\vec{p}) \simeq m\;\;{\rm for \,\; all}\;\, m \in \sigma(H)\,)\;\;\;\Leftrightarrow\;\;\; |\vec{p}| \simeq 0, \label{p=0}
 \end{equation}
and $|\vec{p}| \simeq 0 \;\Leftrightarrow \;$
($|\vec{p}_{\perp}| \simeq 0 $ and $p_{\parallel} \simeq 0$).
Note also that  within the system of units used $|v| < c = 1$. This means
that $|v p_{\parallel}| \leq |v|\, |p_{\parallel}| < |p_{\parallel}| \simeq 0$.
This is why the approximations (\ref{Ekm-a1}) can not be considered as the correct
and consistent with the assumed in \cite{carlo} relation (\ref{Emp-1}).
From the above analysis it follows that the only correct
and self-consistent  approximations are
\begin{eqnarray}
E_{m}(\vec{k}_{m})  \simeq \gamma \,m,\;\;\;
k_{m\,\parallel} \simeq  \gamma v m.
\label{Ekm-b}
\end{eqnarray}
The truth is that
such approximations lead to the result  ${\cal P}_{\vec{v}}(t) = {\cal P}_{0}(\gamma t)$, which was never met in experiments.
So, in the light of the above analysis,
the correctness of the final conclusions drawn in \cite{carlo} is rather questionable.

The another possibility is to assume
that $\vec{p} = const$.
This approach was used
by, e.g. Stefanovich \cite{stefanovich} or Shirokov \cite{shirkov}. It leads to the results
which does not depend on that whether the assumed momentum $\vec{p} = const$ is small or not.
So let us consider now the case of moving quantum system with definite momentum $\vec{p}$. We need the probability amplitude
 $a_{p}(t) = \langle \phi_{p}|\phi_{p} (t)\rangle$, (where $|\phi_{p}\rangle$ corresponds to the moving unstable system with definite momentum $\vec{p}$),
which defines the survival probability
${\cal P}_{p}(t) = |a_{p}(t)|^{2}$.
There is (see \cite{stefanovich,shirkov,khalfin1}),
\begin{eqnarray}
a_{p}(t)
&\equiv &
\int_{\mu_{0}}^{\infty} \omega(\mu)\;
e^{\textstyle{-\,i\sqrt{p^{2}+\mu^{2}}\;\,t}}\,d{\mu}. \label{a-p}
\end{eqnarray}
Results of numerical calculations are presented in Fig (\ref{f-relat-1}),
where calculations were performed for $\omega (\mu) = \omega_{BW}(\mu)$ and $\mu_{0} = 0$, $E_{0}/{\it\Gamma}_{0}
\equiv m_{0}/{\it\Gamma}_{0} = 1000$
and $cp/{\it\Gamma}_{0} \equiv p/{\it\Gamma}_{0} = 1000$.
Values of these
parameters correspond to $\gamma = \sqrt{2}$.
According to the literature
for laboratory systems a typical
value of the ratio $m_{0}/{{\it\Gamma}_{0}}$ is
$m_{0}/{{\it\Gamma}_{0}} \,\geq \, O (10^{3} - 10^{6})$ (see eg. \cite{krauss}) therefore the
choice $m_{0}/{\it\Gamma}_{0} = 1000$ seems to
be reasonable minimum.
Decay curves obtained numerically are presented in Fig  (\ref{f-relat-1}).
\begin{figure}[h!]
\begin{center}
\includegraphics[width=58mm]{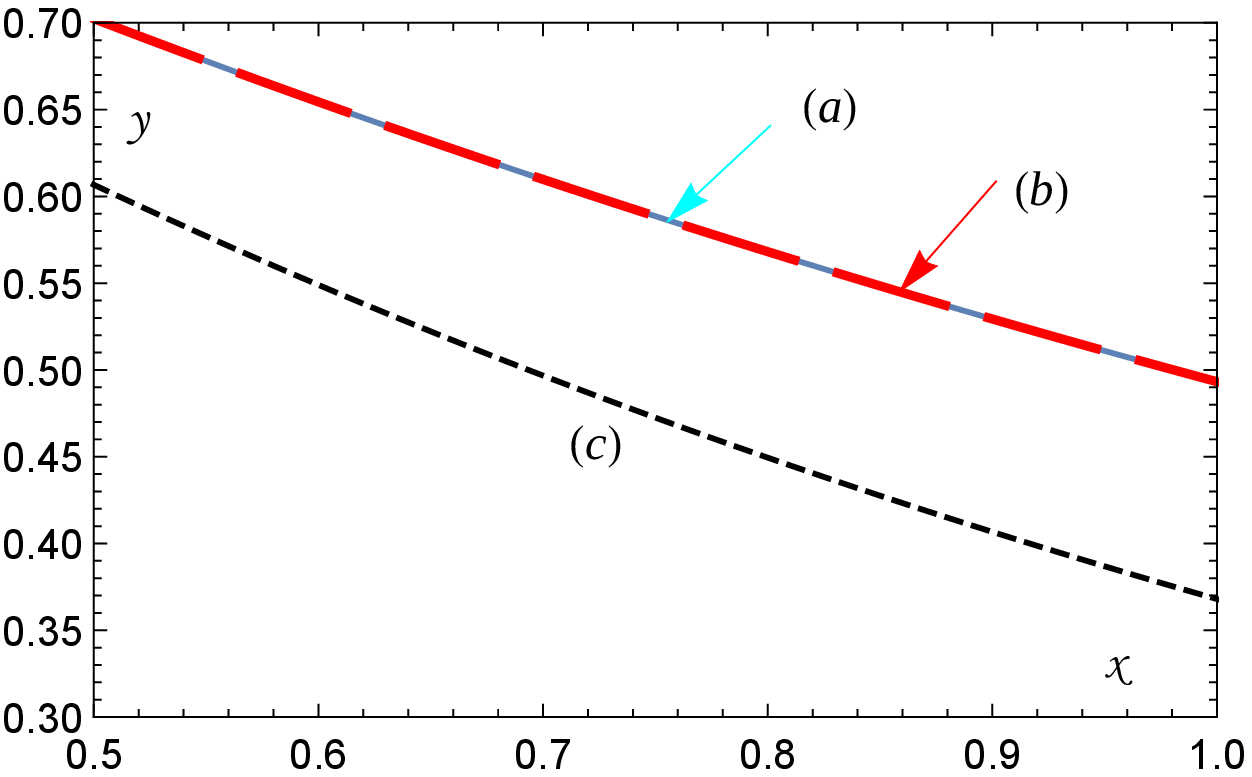} $\;\;\;\;$
\includegraphics[width=58mm]{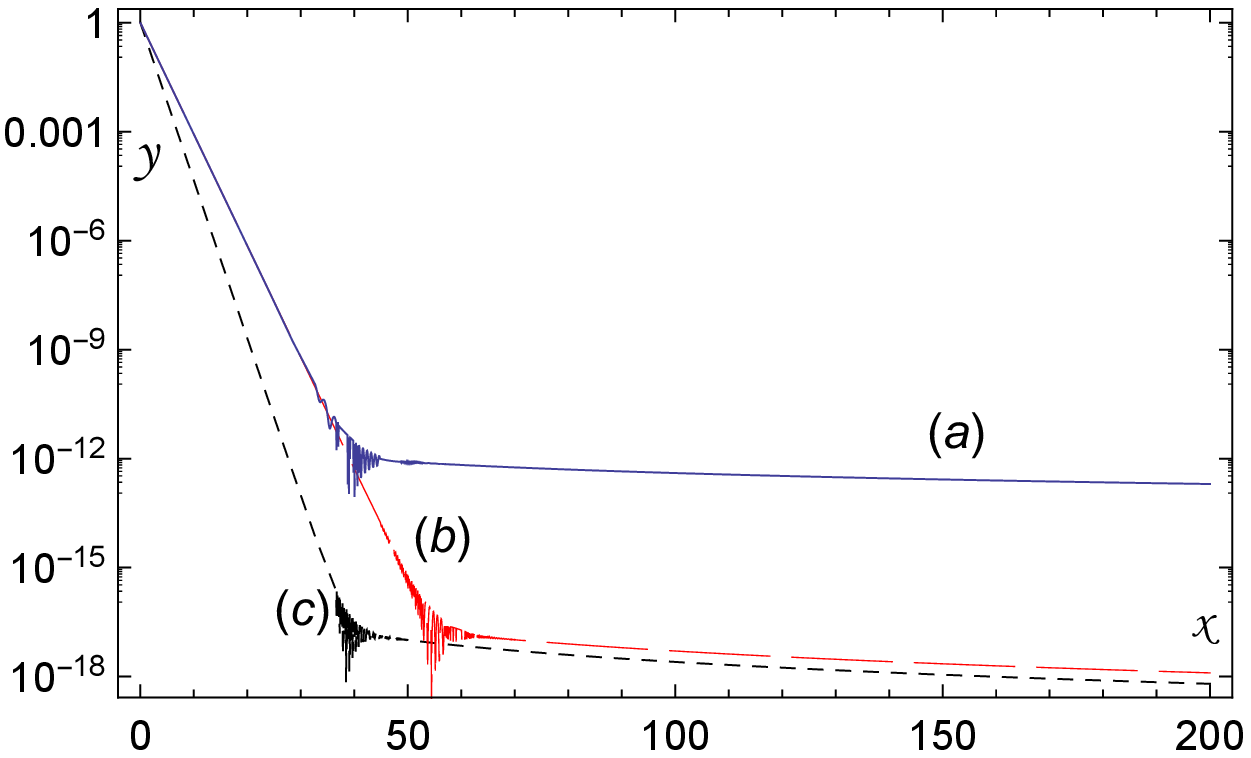}
\caption{ Decay curves obtained for $\omega_{BW}(\mu)$.
Axes: $x =t / \tau_{0} $; $y$ --- survival probabilities:  $(a)$ -- ${\cal P}_{p}(t)$, $(b)$ --
${\cal P}_{0}(t/\gamma)$, $(c)$ -- ${\cal P}_{0}(t)$. }
  \label{f-relat-1}
\end{center}
\end{figure}

Similarly to the case of quantum unstable systems at rest one can calculate the ratio ${\cal P}_{p}(t)/{\cal P}_{c}(t/\gamma)$ in the case of moving particles.
Results of numerical calculations of this ratio are presented in
Figure (\ref{f-Pp-Pc-1})
where
calculations were performed for  $\omega (\mu) = \omega_{BW}(\mu)$ and for $\mu_{0} = 0$,  $m_{0}/{\it\Gamma}_{0} = 1000$, $cp/{\it\Gamma}_{0} \equiv p/{\it\Gamma}_{0} = 1000$ and
$\gamma = \sqrt{2}$.
\begin{figure}[h!]
\begin{center}
\includegraphics[width=65mm]{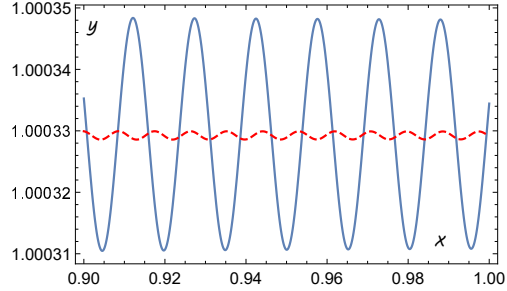}\\
\caption{
Axes: $x =t / \tau_{0} $ --- time $t$ is measured in lifetimes
$\tau_{0}$,   $y$ --- Ratio of probabilities --- Solid line:  ${\cal P}_{p}(t)/{\cal P}_{c}(t/\gamma)$; Dashed line  ${\cal P}_{0}(t/\gamma)/{\cal P}_{c}(t/\gamma)$. }
  \label{f-Pp-Pc-1}
\end{center}
\end{figure}

\section{Summary}
From the results presented in Sec. 3 it follows that
there is no any time interval in which the survival probability (decay) law could be a decreasing function of time of the purely exponential form: In the case of the Breit--Wigner model in so--called "exponential regime" the decay curves are oscillatory modulated with smaller or large amplitude of oscillations depending on the parameters of the model.
In Sec. 4 it it has been shown that in the case of moving relativistic quantum unstable system moving with constant momentum $\vec{p}$, when unstable systems are modeled by the Brei--Wigner mass distribution $\omega(\mu)$, only at times of the order of lifetime $\tau_{0}$
it can be ${\cal P}_{p}(t) \simeq {\cal P}_{0}(t/\gamma)$ to a better or worse approximation. At times longer than a few lifetimes the decay process of moving particles
observed by an observer in his rest system is much slower that it follows from the classical physics relation ${\cal P}_{p}(t) \stackrel{?}{=} \exp\,[- \frac{t}{\gamma}\,{\it\Gamma}_{0}]$: There is ${\cal P}_{p}(t) > {\cal P}_{0}(t/\gamma),\;\;\;{\rm for}\;\;\;t \gg \tau_{0}$ in such a case.
It also appears that in the case of moving relativistic quantum unstable system  with constant momentum $\vec{p}$ decay curves are also oscillatory modulated but the amplitude of these oscillations is higher than in the case of unstable systems at rest. The general conclusion is that
there is a need to test the decay law of moving relativistic unstable system  for times much longer than the lifetime.

\end{document}